\documentclass[fleqn,twoside]{article}
\usepackage{espcrc2}

\usepackage{graphicx}
\usepackage[figuresright]{rotating}

\newcommand{\AmS}{{\protect\the\textfont2
  A\kern-.1667em\lower.5ex\hbox{M}\kern-.125emS}}

\title{The transition from galactic to extragalactic cosmic rays
}

\author{Todor Stanev\address{Bartol Research Institute, 
        Department of Physics and Astronomy, \\ 
        University of Delaware, Newark, DE~19716, U.S.A.}}

\begin{document}

\begin{abstract}
 We discuss the region of transition between galactic and extragalactic 
 cosmic rays. The exact shapes and compositions of these two components
 contains information about important parameters of powerful astrophysical
 sources and the conditions in extragalactic space. Several types of 
 experimental data, including the exact shape of the ultrahigh energy
 cosmic rays, their chemical composition and their anisotropy, and 
 the fluxes of cosmogenic neutrinos have to be included in the 
 solution of this problem. 
\vspace{1pc}
\end{abstract}

\maketitle

\section{INTRODUCTION}

 We believe that at some very high energy of order 10$^{18}$-10$^{19}$ eV
 all observed cosmic rays come from extragalactic sources because 
 they cannot be contained in the Galaxy long enough to be accelerated
 to such high energies~\cite{Cocconi}. The current knowledge of the
 cosmic ray spectrum and composition do not allow us to determine where
 exactly the transition from local (galactic) to extragalactic 
 origin happens. 

  We know that extragalactic cosmic rays lose energy in propagation
 from their sources to us if their sources are isotropically and
 homogeneously distributed in the Universe. The main energy loss
 process is the photoproduction interaction in the microwave background
 (MBR) that  causes the GZK effect, the steepening of the cosmic ray
 spectrum above 6$\times$10$^{19}$ eV~\cite{GZK}. Several fits of the 
 existing experimental data have been published in recent years
 that derived different values of the most important astrophysical
 and cosmological parameters: the acceleration (injection in terms
 of cosmic ray propagation) energy spectrum of these particles and
 the maximum acceleration energy, their chemical composition and
 the cosmological evolution of their astrophysical sources. In the
 assumption that most of ultrahigh energy cosmic rays (UHECR) are
 protons injection spectra as
 different as E$^{-2.7}$~\cite{BGG05} and E$^{-2.0}$~\cite{BW03} and
 cosmological evolution of the type $(1+z)^m$ with $m$ values from
 0 to 3 have been obtained.

 Other attempts~\cite{APO05a,APO05b,HST06} have assumed that extragalactic
 cosmic rays have at their sources the same mixed chemical 
 composition as the low energy galactic cosmic rays. In such a case
 the main energy loss process is the disintegration of the heavy
 nuclei mostly in interactions in MBR. Hadronic interactions
 become important only after the energy per nucleon exceeds the
 photoproduction threshold. The fits of the observed
 cosmic rays spectrum under this assumption gives an intermediate 
 $E^{-2.2-2.4}$ injection spectrum.

 In all these attempts the fits of the observations show the end
 of the galactic cosmic rays spectrum which is obtained by 
 subtraction of the propagated extragalactic spectrum from the
 experimentally observed one. This process gives some limits of
 the astrophysical parameters~\cite{ddmts05} when the subtraction
 gives unphysical negative values. 

 In addition to the cosmic ray spectrum there are several types of
 relevant data: UHECR chemical composition, the anisotropy in this
 energy range, and the production of  {\em cosmogenic}~\cite{BZ69}
 neutrinos by extragalactic cosmic rays. We will briefly discuss 
 these parameters.
 
\section{UHECR ENERGY SPECTRA AND COMPOSITION}

\begin{figure}[htb]
\begin{center}
\centerline{\includegraphics[width=212pt]{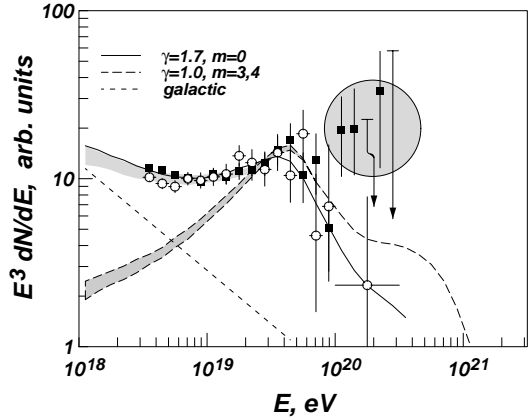}}
\vspace*{-15pt}
\caption{Comparison of two fits of the UHECR spectrum. The solid line
 shows model {\em a} from Ref.~\protect\cite{BGG05} and the dashed
 lines show model {\em b} from Ref.~\protect\cite{BW03}. Short 
 dashes show the galactic component needed to fit the observed
 spectrum in {\em b}.
\label{ssf1a}
}
\end{center}
\vspace*{-40pt}
\end{figure}

 Figure~\ref{ssf1a} compares two different fits of the extragalactic
 cosmic ray spectra in the assumption that they are purely protons
 and that their differential acceleration spectrum is a power law
 $E^{-{\gamma+1}}$.
 The experimental data are from AGASA~\cite{AGASA} and HiRes~\cite{HiRes}
 and are normalized to each other at 10$^{19}$ eV. Since we are now
 only interested in the shape of the spectrum the exact differential
 flux at the normalization points is not important. It is obvious
 that the two experimental measurements agree well with each other
 on the shape of the energy spectrum with exception of the AGASA events
 above 10$^{20}$ eV. The most recent analysis of the AGASA data, presented
 this summer~\cite{Teshima-here} decreases the energy assignment 
 of the AGASA data by 10-15\% and makes the spectrum closer to that
 of HiRes.

 Fit {\em a}~\cite{BGG05} derives an injection spectrum
 with $\gamma$ = 1.7. The dip at about 10$^{19}$ eV is due to the
 transition of
 proton energy loss to Bethe-Heitler $e^+e^-$ pair production
 to purely adiabatic loss 
 as predicted by Berezinsky\&Grigorieva~\cite{BG88}. The model 
 does not need any contribution from galactic cosmic rays to
 describe the observed cosmic ray spectrum down to 10$^{18}$ eV.
 There is also no need for cosmological evolution of the extragalactic
 cosmic ray sources, although some source evolution can be
 accommodated with a slight change of $\gamma$. The model predicts
 purely proton composition of the extragalactic cosmic rays
 and does not work as well as shown in Fig.~\ref{ssf1a} 
 if more than about 10\% of the cosmic rays at the source are 
 nuclei heavier than H.
 Fit {\em b}~\cite{BW03} does need contribution from galactic
 sources with $E^{-3.50}$ spectrum that extends well above
 10$^{19}$~eV. The extragalactic contribution is shown for two
 different cosmological evolutions with $m$ = 3 (as used in Ref.~\cite{BW03})
 and 4 (upper edge of the shaded area). The influence of the
 cosmological evolution on the cosmic ray propagation is modest
 because no protons injected at redshifts $z$ higher than 0.4 arrive
 at Earth with energy above 10$^{19}$~eV independently of their
 initial energy.

 Obviously these two models predict very differently the end of
 the galactic cosmic ray spectrum. In model {\em a} the galactic
 cosmic ray sources do not need to accelerate particles above
 10$^{18}$~eV. In model {\em b} they should be able to reach 
 energies higher by one and a half orders of magnitude. This would 
 affect very strongly the expected cosmic ray composition.

 The assumption that extragalactic cosmic rays have a mixed 
 composition at acceleration gives somewhat intermediate results
 for the injection spectrum of UHECR~\cite{APO05a,APO05b,HST06}.
 The spectrum that fits the observation best has $\gamma$ values
 between 1.2 and 1.4. The chemical composition of cosmic rays 
 at Earth is also different and obviously depends on the 
 source composition.   
 
 It used to be that we considered three separate regions in the
 cosmic ray spectrum - energy below 10$^{15}$ eV where the
 acceleration is at supernova remnants, 10$^{15}$-10$^{19}$ eV
 where acceleration is  at {\em unknown} galactic sources and   
 extragalactic component. Since stochastic shock acceleration 
 is rigidity dependent (as is the escape from the Galaxy) this
 picture predicts a complicated composition energy dependence.
 When a source cannot accelerate protons any more, there are
 only heavier nuclei, initially He, then CNO, then Si and
 finally Fe nuclei that are accelerated. 
 One may imagine the composition becoming heavier between
 10$^{15}$ and 10$^{16}$ eV, then lighter at higher energy when
 the {\em unknown} sources take over, then heavier again when they
 are exhausted, then lighter again when extragalactic cosmic
 rays start dominating. Recent developments in the acceleration
 theory~\cite{TBell} show that the maximum acceleration energy
 may be much higher~\cite{Blasi06} and galactic sources can cover 
 the whole energy range to 10$^{18}$-10$^{19}$ eV, i.e.
 we expect one single transition region. The Kascade
 experiment has already observed~\cite{Kascade}
 the beginning of the exhaustion of the galactic 
 cosmic ray sources and the respective change of the composition. 

\begin{figure}[htb]
\begin{center}
\vspace*{-5truemm}
\includegraphics[width=212pt]{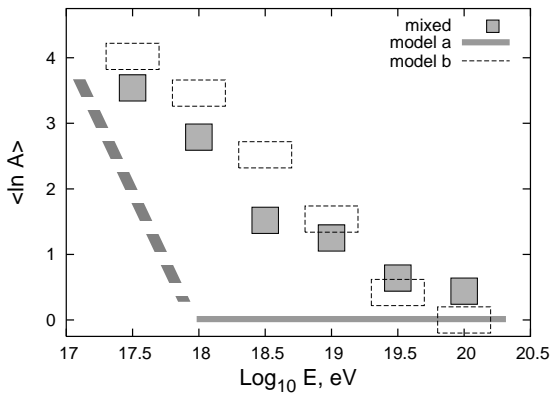}
\vspace*{-7truemm}
\caption{Cosmic ray chemical composition as a function of the 
 total energy per nucleus in the three models. 
 The mixed composition model data are taken 
 Ref.~\protect\cite{APO05b}. The error bars assumed are
 logarithmic and probably lower than the true experimental
 ones. The very heavy dashed line represents symbolically the
 results of the HiRes prototype/MIA coincidences.  
\label{composi}
}
\end{center}
\vspace*{-10truemm}
\end{figure}

 Figure~\ref{composi} shows the
 predicted composition as a function of the total energy per 
 nucleus in the transition region in the classical cosmic ray
 units of $<ln A>$. This is very appropriate when the detection
 is through air showers with logarithmic sensitivity to both
 energy and composition.
 Model {\em a} presents the easiest case to explain. All composition 
 changes happen below 10$^{18}$~eV. We do not plot the composition below
 that energy as it is not very well defined. The results for the other
 two models are presented using our own understanding of them.
 For model {\em b} we assume that at 10$^{17.5}$~eV all cosmic rays
 are iron and assign $<ln A>$ value of 4. At higher energy we use 
 the fraction of galactic cosmic rays to iron nuclei and thus calculate
 the corresponding $<ln A>$ value. For the mixed composition we use 
 Fig.~3 of Ref.~\cite{APO05b} where the cosmic ray composition at
 Earth is plotted for $\gamma$=1.3 and $E_{max}^Z$ is given as 
 $Z\times$10$^{20.5}$~eV. 

 Surprisingly the difference in the energy dependence of $<ln A>$ 
 of model {\em b} and of the mixed composition model is not that big,
 although in the model {\em b} case the composition only consists of
 a combination of Fe and H, and in the mixed composition case we
 have five groups of nuclei.

 Model {\em a} gives a very different picture, at least above
 10$^{18}$~eV, where the composition is purely Hydrogen. It is 
 definitely distinguishable from the other two models. The
 prediction that the composition does not change above 10$^{18}$~eV
 and is very light is supported by the HiRes $X_{max}$
 measurement~\cite{HiRes_comp}, indicated with heavy dashes in
 Fig.~\ref{composi}. What is the exact meaning of
 the detected {\em light} composition is not known because of
 the differences between the hadronic interaction models used
 for data analysis.
  
 On the other hand, other experiments support a much milder
 energy dependence of the cosmic ray chemical composition, that is 
 more in line with the prediction of of models {\em b} and that
 of mixed extragalactic cosmic ray composition.

\section{ANISOTROPY}

 Low energy cosmic rays diffuse in the magnetic fields of the Galaxy
 and lose memory of the location of their sources. The anisotropies
 are very small, well below 1\% and are difficult to measure.
 The measurement of a small anisotropy with air shower experiments
 requires an exact knowledge of the experimental acceptance and of
 the lifetime of the shower array. 

 At higher energy things are supposed to change as the cosmic ray
 rigidity increases and their diffusion in the Galaxy becomes faster.
 The first question we still cannot answer is of the rigidity
 (energy) dependence of the cosmic ray diffusion coefficient.
 An energy dependence of $E^{-0.5-0.6}$ is derived from the ratio of 
 secondary to primary cosmic rays. If this dependence is extended
 by seven orders of magnitude then galactic UHE protons should
 show very strong anisotropy that is not observed. The two spots
 with increased cosmic ray flux identified by the AGASA group~\cite{A_aniso}
 (one in the general direction of the Galactic center and one in the
 direction of Orion) are not confirmed by other experiments.
 Theoretically we expect Kolmogorov turbulence in the Galaxy, that
 should under some circumstances give $E^{1/3}$ energy dependence.

 One way to explain the low anisotropy in the beginning of the 
 transition region is to assume that all galactic cosmic rays are
 heavy nuclei. If they were indeed heavy nuclei
 the average particle rigidity would 
 significantly decrease and would maintain the high isotropy
 of the galactic cosmic rays. In the transition region, however,
 the cosmic ray chemical composition becomes lighter and 
 correspondingly we expect to see some degree of anisotropy.

 I am convinced that in the case of high experimental statistics
 (as expected from the Pierre Auger Observatory~\cite{Auger})
 the anisotropy in the transition region between galactic and
 extragalactic cosmic rays deserves a careful study. A part of
 it is theoretical. We should propagate particles of different
 rigidity in the Galaxy and attempt to understand their general
 behavior. Analytic solutions of the diffusion equation are not
 any more suitable for this problem. Particle trajectory has to be 
 numerically solved, most likely in a Monte Carlo fashion, and in
 detailed enough magnetic field models. Such models should be
 tested to match analytic calculations when applied to the appropriate
 simplified models.
 
\section{COSMOGENIC NEUTRINOS}

 Cosmogenic neutrinos are produced in the same photoproduction
 interactions of the UHECR protons that create the GZK effect.
 They were first proposed by Berezinsky\&Zatsepin in 1969~\cite{BZ69}
 and were the subject of many calculations afterward.
 Cosmogenic neutrinos are often considered a `guaranteed source'
 of ultrahigh energy neutrinos. They are indeed guaranteed, since
 we know UHECR exist, but their flux is uncertain.

 An essential quality of neutrinos is that they have a low 
 interaction cross section. This makes neutrino detection a
 difficult problem that requires huge detectors of at least
 km$^3$ scale. Such detectors are now in the stage of
 planning~\cite{KM3Net} and construction~\cite{IceCube,ANITA}.
 The question of the cosmogenic neutrino flux and of its 
 relation to other data from ultra-high astrophysics experiment
 is very timely.

 The low interaction cross section is not only a deficiency.
 While protons emitted at redshifts $z$ exceeding 0.4 do not
 reach us with energy above 10$^{19}$~eV the cosmogenic 
 neutrino production peaks at redshifts exceeding 2 for 
 $(1+z)^3$ cosmological evolution of the UHECR sources.
 This is the main link between the extragalactic UHECR
 and cosmogenic neutrinos. The detection of cosmogenic 
 neutrinos may help the degeneracy in modeling of the
 extragalactic cosmic rays spectra shown in Fig.~\ref{ssf1a}.

\begin{figure}[htb]
\begin{center}
\includegraphics[width=212pt]{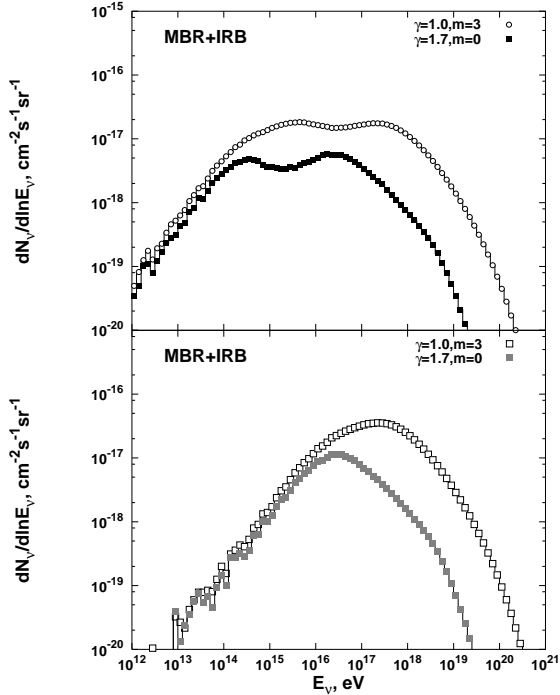}
\vspace*{-8truemm}
\caption{Fluxes of cosmogenic neutrinos generated in proton interactions
 in the isotropic photon background~\protect\cite{SDMMS06} in the two models
 of the UHECR spectrum. The upper panel compares the fluxes of 
 \protect$\nu_e + \bar{\nu}_e$. The lower one compares the fluxes of
 \protect$\nu_\mu + \bar{\nu}_\mu$.
\label{all_nu}
}
\end{center}
\vspace*{-5truemm}
\end{figure}

 The point is that models that use flat cosmic ray injection
 spectrum and require strong cosmological evolution of the
 UHECR sources, such as model {\em b} would produce significantly
 more cosmogenic neutrinos than steep injection spectrum models
 with no cosmological evolution. Such a comparison between the
 two models is shown in Figure~\ref{all_nu}.
 The figure shows cosmogenic neutrino fluxes generated by proton 
 interactions in MBR~\cite{ESS01} and in the infrared/optical
 background (IRB)~\cite{SDMMS06}
 with the injection spectrum and cosmological
 evolution of the two models as indicated in the two panels.
 The injection spectra of the two models are normalized to 
 the UHECR differential flux at 10$^{19}$ eV.

 Before discussing the magnitude of the two fluxes I want to 
 attract your attention to the flux of electron neutrinos and
 antineutrinos that exhibits two maxima separated by about two orders 
 of magnitude of energy. The higher energy one is due to the
 $\pi^\pm \rightarrow \nu_\mu + \bar{\nu}_\mu + \nu_e$ decay final
 state and consists mostly of electron neutrinos. It peaks at
 the same energy where $\mu_\mu$ and $\bar{\nu}_\mu$ do. The lower 
 energy maximum is due to electron antineutrinos from neutron
 decay. Comparison between the neutron interaction length
 and their decay length show that all neutrons of energy below 
 10$^{20.5}$~eV decay rather than interact.

 The number density of IRB is of order 1 and varies by about
 a factor of 2 in different models, but its energy extends to
 energies exceeding 1 eV. What that means is that extragalactic
 particles of much lower energy would interact in IRB and will
 generate neutrinos.
 Reference~\cite{SDMMS06} shows the neutrino yields generated by
 protons of energy as low as 10$^{18}$~eV. This yield is small but
 has to be weighted by the much higher number of particles at that
 energy - almost 1,000 higher that that of 3$\times$10$^{19}$~eV 
 for a flat $E^{-2}$ injection spectrum. A more recent calculation
 that also includes UV photons~\cite{AAB06} employs 
 interactions of 10$^{17}$ eV nucleons that have to be weighted
 still higher. The cosmological evolution of IRB is not as strong
 as that of MBR, as the radiation is generated through emission
 directly by stars and after absorption by dust, but still significant
 up to redshift of 3. The IRB model of Ref.~\cite{SMS} is used for
 the shown calculation. 

 Model {\em b} generates much higher cosmogenic neutrino fluxes
 than model {\em a} because of two reasons that contribute
 roughly the same increase of the cosmological neutrinos.
 Firstly, it uses much flatter injection spectrum $E^{-2.0}$
 which means equal amount of energy per decade. It thus
 contains many more particles above the photoproduction
 interaction threshold in the MBR is about
 3$\times$10$^{19}$~eV and, of course, decreases as $(1+z)^{-1}$.

 The other reason is that model {\em b} employs a strong
 cosmological evolution of the cosmic ray sources. This
 increases by $\sqrt{3}$ the number of particles injected
 at redshift of 2, but in addition increases by a factor
 of three the number of particles above the interaction threshold.
 This way the total neutron flux is increased by the cosmological
 evolution of the sources by a factor of five. 

 The difference in the peak values of the cosmological neutrinos
 generated by the two models is more than one order of magnitude.
 In practical terms, after the energy dependence of the $\nu N$ 
 cross section is taken into account, this means that model {\em b}
 generates  fluxes that are in principle detectable by IceCube at
 the rate of roughly less than one event per year, while model {\em a} 
 generates undetectable fluxes of cosmogenic neutrinos in km$^3$
 detectors.

  Ice and water neutrino detectors are generally not
 very suitable for cosmogenic neutrino detection. Much better 
 strategies for these UHE neutrinos are the radio and acoustic
 detectors that have very high detection threshold, but also
 will have higher effective volume. The other option are
 giant air shower arrays such as Auger, that can reach 
 effective volume of 30 km$^3$ and, with sufficiently low
 threshold (10$^{18}$~eV) could detect several events per 
 year. 

 Cosmogenic neutrinos are also generated in the mixed composition
 scenario~\cite{HTS05,ABO05}. Since the major energy loss
 process is the disintegration of heavy nuclei, the main flux 
 component of $\bar{\nu}_e$ from neutron decay. The 
 $\bar{\nu}_e$ flux exceeds by a factor of 5 the sum of the 
 neutrino fluxes of all other flavors in Ref.~\cite{ABO05}.
 The absolute magnitude depends again mainly on the injection spectrum
 and the cosmological evolution of the sources.\vspace*{-2truemm}
  
\section{DISCUSSION}

 The transition region between galactic and extragalactic cosmic 
 rays is currently not very well defined and studied. It has
 a significant astrophysical importance, because the energy content
 of the extragalactic cosmic rays, as well as their injection
 spectrum and composition, will define much better the type of 
 their sources. In addition, a derivation of the cosmological
 evolution of the sources will not only restrict the number of
 available scenarios, but also provide an additional measure
 of the cosmological evolution of powerful astrophysical objects.

 The first experimental result on the shape of this transition
 presented by the Fly's Eye group~\cite{Birdetal93} used the
 simultaneous change of the cosmic ray composition and energy
 spectrum shape. We should now employ all available types of
 measurements to solve this problem. These measurements 
 include the cosmic ray composition in the transition region 
 and above it, the cosmic ray anisotropies, and the possible
 detection of cosmogenic neutrinos.

 The possible detection of cosmogenic neutrinos would be a
 powerful test if we succeed in collecting a reasonable statistics
 of such events. It is unlikely this will happen very soon, but
 the expanding efforts for designing and building detectors for
 ultrahigh energy neutrinos are very encouraging.

 {\bf Acknowledgments} This talk is based on work performed in 
 collaboration with D.~Seckel, D.~DeMarco, J.~Alvarez-Mu\~{n}iz,
 R.~Engel and others. Partial support of my work comes from 
 US DOE Contract DE-FG02 91ER 40626 and NASA Grant ATP03-0000-0080.

\end{document}